\documentclass[conference, a4paper]{IEEEtran-modified}

\IEEEoverridecommandlockouts

\usepackage{cite}       

\ifx\pdfoutput\undefined
	\usepackage{graphicx}
\else
	\usepackage[pdftex]{graphicx}
\fi

\usepackage{subfigure} 

\usepackage{url}       

\usepackage{amsmath}    
\usepackage[latin1]{inputenc}   
\usepackage[T1]{fontenc}   
\renewcommand{\keywordname}{Keywords}

\usepackage{algorithm}
\usepackage[noend]{algpseudocode}

\begin{document}

\graphicspath{{img/}}


\title{Association rule mining and itemset-correlation based variants}


\author{
\authorblockN{Niels M\"undler}
\authorblockA{Department of Informatics\\Technische Universit\"at M\"unchen\\
Email: n.muendler@tum.de} 
}


\specialpapernotice{Seminar Data Mining}


\maketitle




\begin{abstract}
    Association rules express implication formed relations among attributes in databases of itemsets.
    The apriori algorithm is presented, the basis for most association rule mining algorithms.
    It works by pruning away rules that need not be evaluated based on the user specified minimum support
    confidence.
    Additionally, variations of the algorithm are presented that enable it to
    handle quantitative attributes and to extract rules about generalizations of items,
    but preserve the downward closure property that enables pruning.
    Intertransformation of the extensions is proposed for special cases.
\end{abstract}


\begin{keywords}
Data Mining Quantitative Generalized Association Rule Mining
\end{keywords}


\section{Introduction}

First introduced by Agrawal et al. in \cite{DBLP:conf/sigmod/AgrawalIS93} as an extension for existing databases,
association rules provide a means for discovering in a large database of items that appear together
implications of the form "if $\text{item}_1, \text{item}_2, \dots $ are in the set then also $\text{item}_k, \text{item}_{k+1}, \dots$ are in the set"
associated with a measure for the probability that this implication holds.
A first application domain for this emerged in the area of shopping
where digitalization made large amounts of such data available.
Through extraction of association rules an insight on consumer behaviour should be gained.

The database contains a set of transactions which contain all of the items bought by a customer at once.
An association rule $\{Aubergine, Charcoal\} \rightarrow \{Beer\}$ means that when customers
bought aubergines and charcoal, they also often bought beer.
Buying beer though does not have to imply that either aubergines or charcoal are bought, for example when drunk with weisswurst for breakfast.
Thus not all association rules are symmetrical.
This rule is said to have support of 10\% if aubergine, charcoal and beer were contained in 10\% of all transactions.
The percentage of transactions that also contained beer when aubergine and charcoal were contained is called confidence.
Usually a user-specified minimum support and minimum confidence for extracted rules is specified.

It can easily be seen that a data base with $n$ different items
there are $2^n$ possible association rules.
Hence, based on the minimum support and confidence, sensible pruning mechanisms have to be used such that
not many more rules are evaluated than are included in the result set.
In the pioneer works of Agrawal et al. \cite{DBLP:conf/sigmod/AgrawalIS93,DBLP:conf/vldb/AgrawalS94} algorithms that perform well on large datasets
are proposed, among them the apriori algorithm which will be explained in detail in \autoref{chap:apriori}.
In addition, common variations of the apriori algorithm are presented that make it possible
to work on datbases with quantitative data and with generalizations of the items. 
All of the presented variations preserve the downward closure property of itemsets that are to be generated,
making it possible to use the main pruning strategy of the apriori algorithm.

For related work, a very broad overview over the topic of data mining in general in databases is given by Chen et al. in
\cite{DBLP:journals/tkde/ChenHY96}, yet focusing not too much on association rules.

\section{Association Rules}

\subsection{Motivation}

Consider the database of a supermarket.
The management of the supermarket might be interested in which items appear often together
in the shopping baskets of their customers.
This information can then be used for strategic decisions.
For example if the market knows that $\{Aubergine\} \rightarrow \{Charcoal\}$
when providing more aubergines to the customers, more charcoal should be provided too.
Or if all rules of the form $X \rightarrow Beer$ were known,
the sale of beer could be boosted by placing it near to items in $X$ or
by reducing the price of the items in $X$.
Of course the management is only interested in behavior of a significant amount
of customers and implications that hold for a large proportion of the transactions where the left side is satisfied.
In the following sections, a solution to this problem is described that was introduced by Agrawal and Srikant in \cite{DBLP:conf/vldb/AgrawalS94}.

\subsection{Formal definition}
\label{chap:formal_crisp_ass_rules}

The definition is based on the definition introduced in \cite{DBLP:conf/sigmod/AgrawalIS93}.
For a set of attributes $A$, an association rule is a rule of the form $X \rightarrow Y$ where $X, Y \subseteq A^{+}$ and $X \cap Y = \emptyset$.
$X$ is called the antecedent and $Y$ the consequent of the rule and the elements of those sets are called items.
Sets of $k$ items are also called $k$-itemsets.
An association rule $X \rightarrow Y$ is said to be contained in a transaction or itemset $T=\{t_1, ..., t_n\} \subseteq A^n$ if $X \cup Y \subseteq T$.
Similarly an itemset $I$ is contained in $T$ if $I \subseteq T$.
The database or dataset $D$ is the set of all collected transactions.
A rule or itemset $I$ has $support(I) := s\%$ if it is contained in $s\%$ of the transactions in the database.
This can be used as a sign of statistical significance.
Also, a rule $X \rightarrow Y$ has $confidence(X \rightarrow Y) := c\%$ if for $c\%$ of the transactions $T$ with $X \subseteq T$ also holds $Y \subseteq T$,
which means that the rule is contained in $c\%$ of the transactions that do contain the antecedent.
It can be regarded as equivalent to $Pr_D [Y|X]$, the likelihood of $Y$ also "occuring" when $X$ is given, based on the database $D$.

Usually there is a user defined minimum confidence and minimum support,
such that all extracted association rules have a support of at least the minimum support and
a confidence of at least the minimum confidence.

An itemset that has at least the minimally specified support is called a frequent itemset.
An arbitrary total order on the attributes in the database is introduced,
and all itemsets and transactions are regarded as tuples ordered with respect to this order.

\subsection{Problem decomposition}

In the process of extracting all association rules that do have minimum support and minimum confidence,
an algorithm must
\begin{itemize}
    \item Generate frequent itemsets $X$
    \item Evaluate all association rules $X - Y \rightarrow Y$ where $Y \subset X$ and keep those that satisfy minimum confidence and support
\end{itemize}

It suffices to generate frequent itemsets because all of the corresponding association rules have the same support
and we are only interested in association rules which have at least minimum support.
The apriori algorithm presents an efficient method for the generation of frequent itemsets
by only considering combinations of smaller frequent itemsets. It is described in detail in \autoref{chap:apriori}.
A method for the efficient generation of association rules from the frequent itemsets is described in \autoref{chap:discovering_rules}.

\subsection{The Apriori Algorithm}
\label{chap:apriori}

\begin{figure*}[t]
    \centering
    \subfigure[Hasse diagram, layered by size.]{\includegraphics[width=0.22\linewidth]{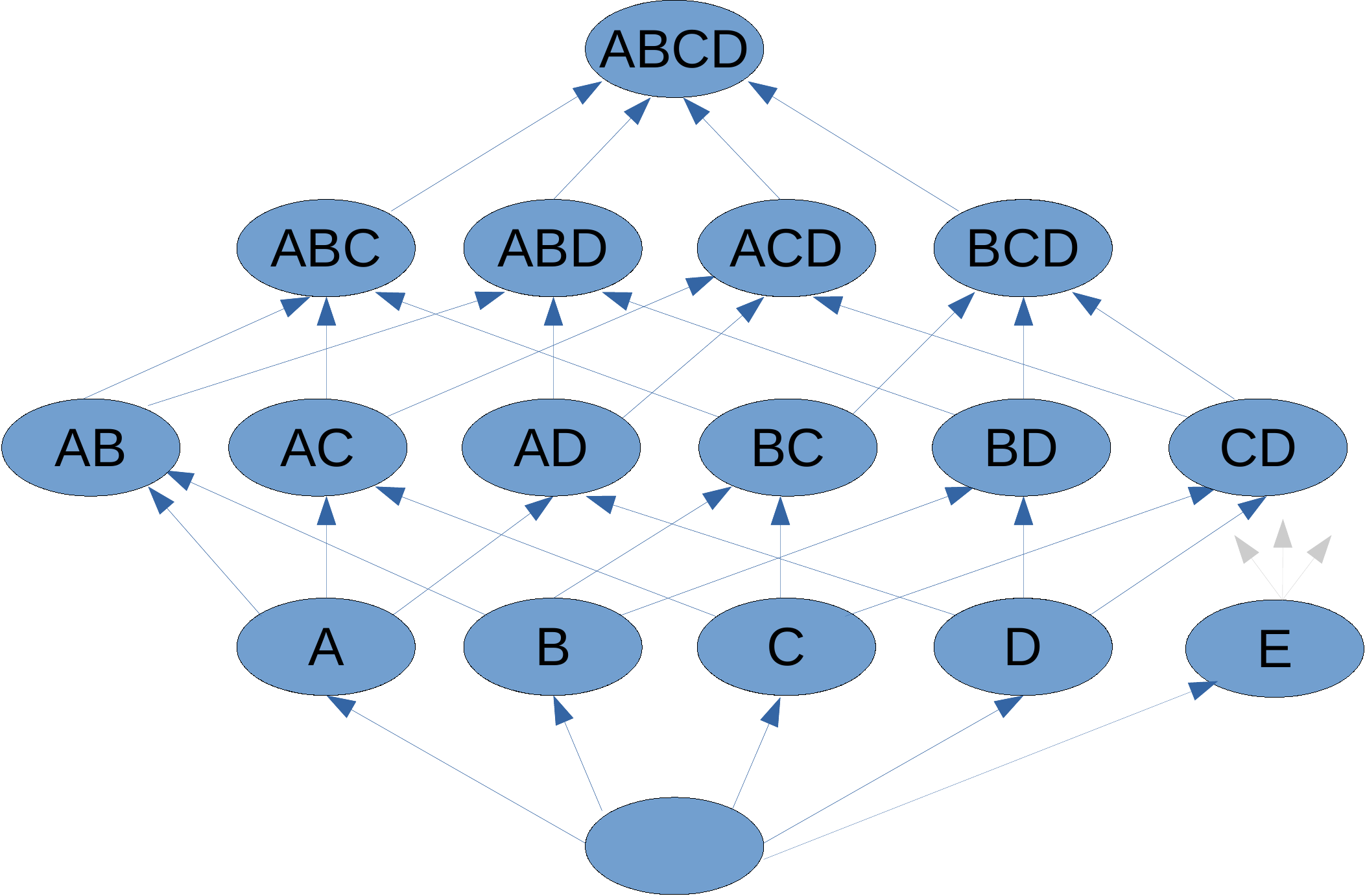}}%
    \qquad
    \subfigure[Join-constraint edges, $k=1$]{\includegraphics[width=0.22\linewidth]{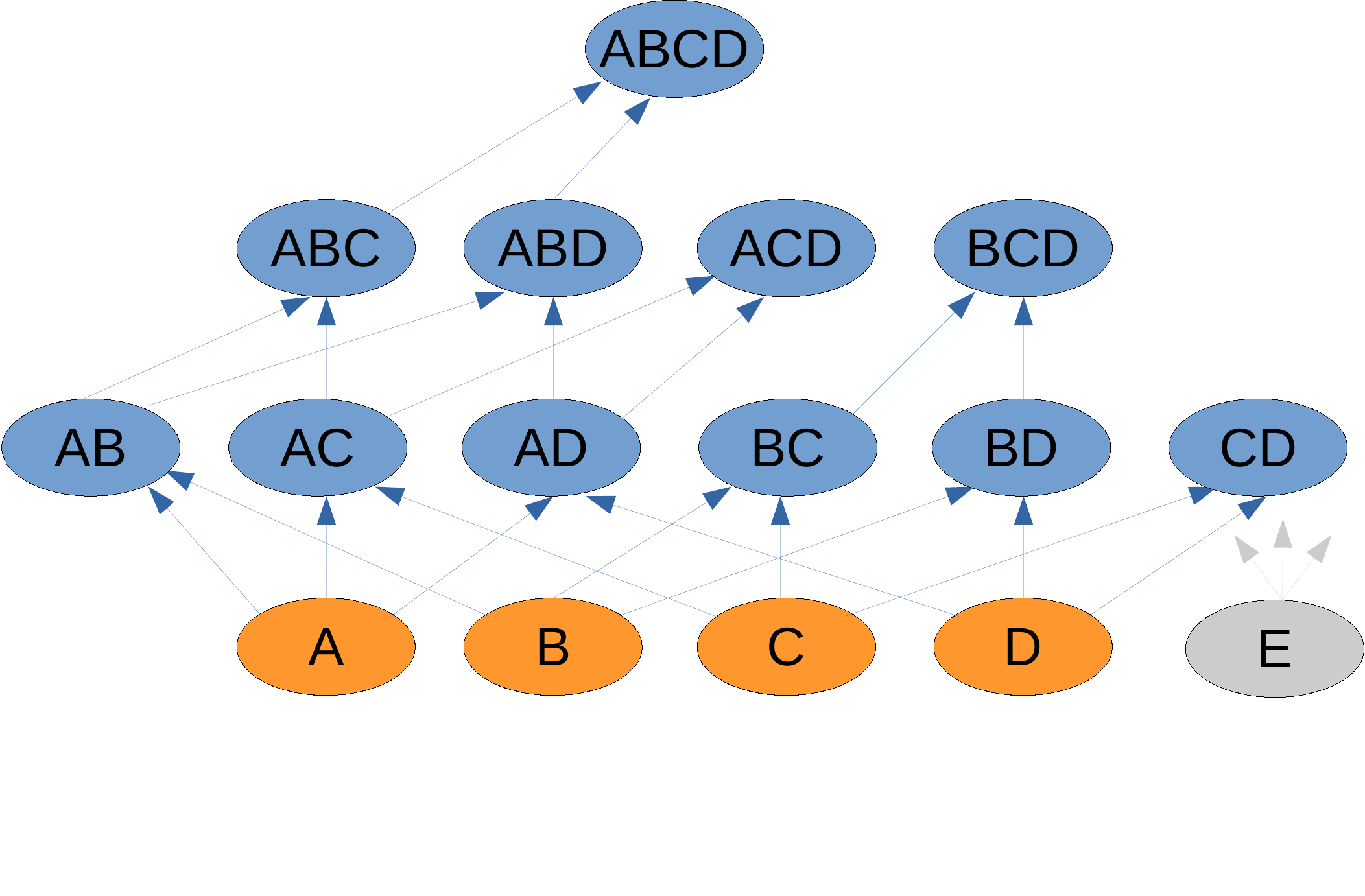}}%
    \qquad
    \subfigure[$k = 2$]{\includegraphics[width=0.22\linewidth]{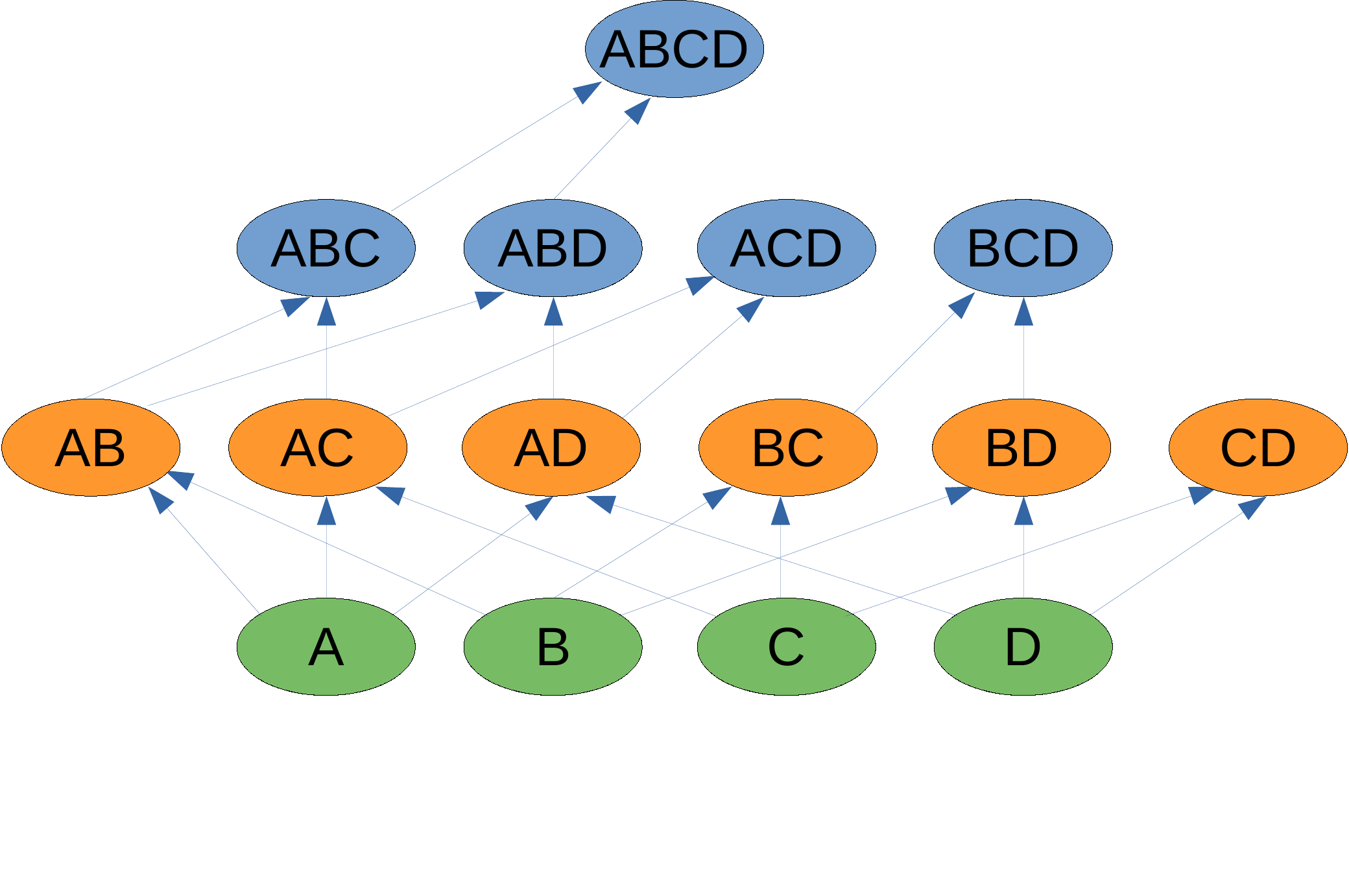}}%
    \qquad
    \subfigure[$k = 3$]{\includegraphics[width=0.22\linewidth]{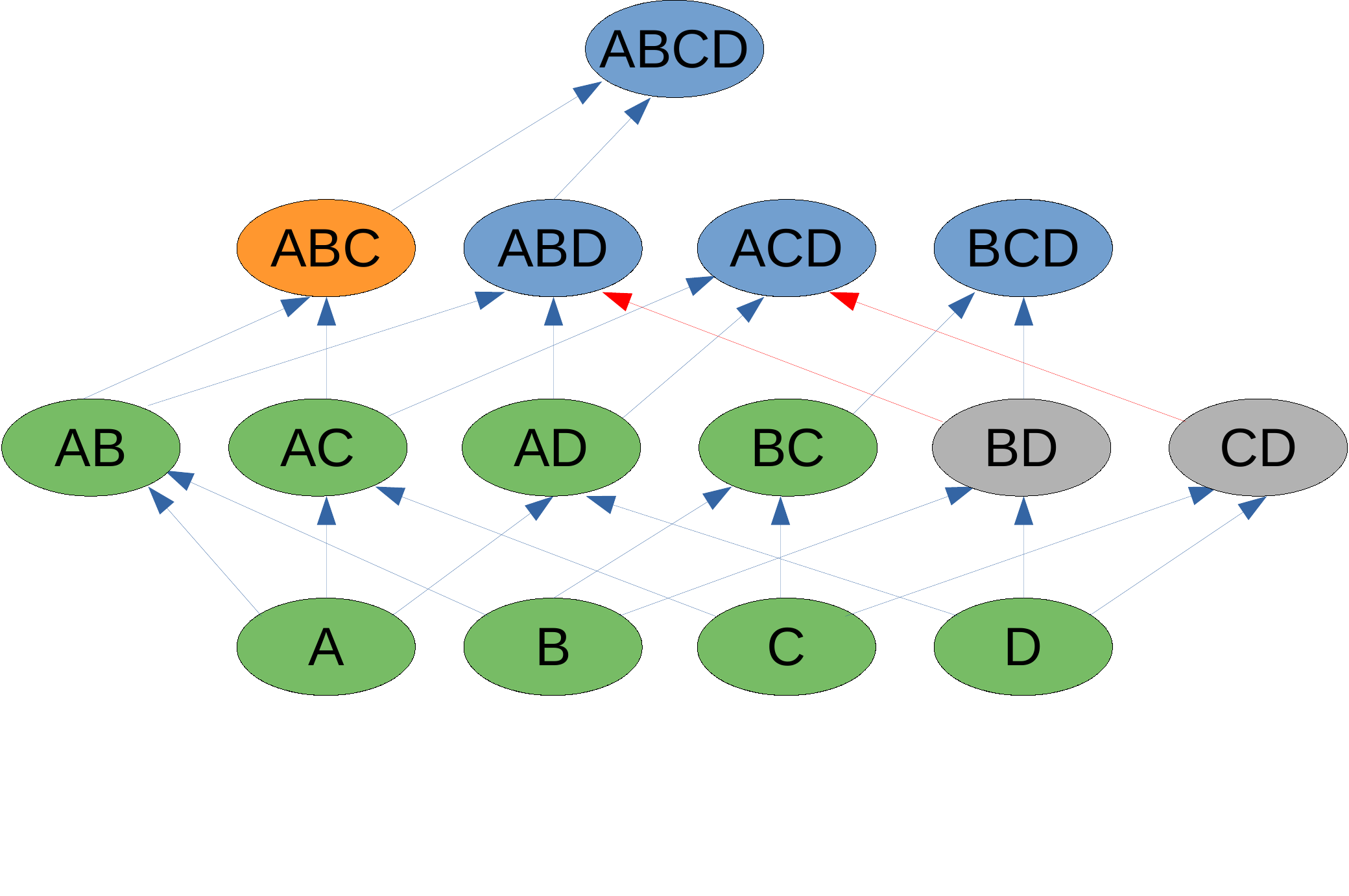}}%
    \caption{Visualization of the frequent itemset generation of the apriori algorithm on $ABCDE$.
        Green vertices have been identified as frequent itemsets. Candidate itemsets are orange. Blue nodes are never checked.
        In the last step some vertices are pruned as they contain the non-frequent subsets $BD, CD$ (red edges).}
    \label{fig:visualization-generation}
\end{figure*}

The approach is based on the observation that every subset of an itemset has to have at least the same support.
This can be seen easily as every subset of the itemset $I$ is also contained in the transaction that originally contained $I$.
It follows that if any itemset $I$ is not frequent, all larger sets that contain $I$ are also not frequent.
Thus, for generating candidate frequent itemsets of size $k+1$ it suffices to consider candidate itemsets of size $k+1$ that are unions of frequent itemsets of size $k$.
For each of the candidates, the actual support in the database is checked by scanning the whole database.
After each scan, the actual frequent itemsets are used for the next iteration.
The overall procedure can be seen in \autoref{fig:visualization-generation} and the algorithm is shown in \autoref{alg:apriori}.

\begin{algorithm}
 \caption{Apriori Frequent Itemset Generation from \cite{DBLP:conf/vldb/AgrawalS94}}
 \label{alg:apriori}
 \begin{algorithmic}[1]
 \Function{Apriori}{database of transactions $D$}
  \State $L_1 \gets $\{frequent 1-itemsets\}
  \For {$k \gets 2; L_k \neq \emptyset ; k++$}
    \State $C_k \gets$ apriori-gen($L_{k-1}$) 
    \ForAll {transactions $t \in D$}
      \State $C_t \gets$ subset($C_k , t$) \Comment{Candidates $\subseteq t$}
      \ForAll {candidates $c \in C_t$}
        \State c.count++
      \EndFor
    \EndFor
    \State $L_k \gets \{c \in C_k | c.count \geq minimum support\}$
  \EndFor
  \State \Return $\bigcup_k{L_k}$ \Comment{All frequent itemsets in $D$}
 \EndFunction
\end{algorithmic}
\end{algorithm}
 
\subsubsection{Candidate generation}
\label{chap:candidate_generation}
In order not to generate any itemset multiple times,
only $k$-itemsets are combined into a $k+1$-itemset where the first $k-1$ items are equal.
This results in one unique way to construct a set from smaller sets.
For example ABCD will only be constructed from ABC and ABD as all other combinations of $3$-itemsets already differ in the first or second item.
Additionally this ensures that the result is maximally of size $k+1$.
Hence in the join phase of \autoref{alg:candidate_generation} candidate itemsets of size $k+1$ are generated by a join of the frequent itemsets $F_k$ of size $k$
on the condition of being equal in the first $k-1$ items and not being equal for the last item.

Assuming that all generated frequent sets size $k$ were already generated,
due to the above observation if any subset $I_k \subset I_{k+1}$ is not among the already generated sets, $I_k$ has to be non-frequent.
Then, $I_{k+1}$ is non-frequent too.
Thus in the prune step of \autoref{alg:candidate_generation} it is checked if all $k+1$ $k$-subsets of a newly generated itemset were already generated. 

\begin{algorithm}
  \caption{Generation of candidate frequent $k$-itemsets from frequent $k-1$-itemsets from \cite{DBLP:conf/vldb/AgrawalS94}}
  \label{alg:candidate_generation}
\begin{algorithmic}[1]
\Function{apriori-gen}{$L_{k-1}$}
    \State \textbf{insert into} $C_k$ \Comment{Join}
    \State \textbf{select} $a.item_1, \dots, a.item_{k-2}, a.item_{k-1}, b.item_{k-1}$
    \State \textbf{from} $L_{k-1}\ a, L_{k-1}\ b$
    \State \textbf{where} $\forall{i \in [1, k-2]: a.item_i = b.item_i}$
        \State \textbf{and} $a.item_k \neq b.item_k$
    \ForAll{$c \in C_k$} \Comment{Prune}
       \ForAll{$k-1$ subsets $s \subset c$} 
          \If {$s \notin L_{k-1}$}
            \State \textbf{delete} $c$ from $C_k$
            \State \textbf{continue}
            \EndIf
        \EndFor
    \EndFor
    \State \textbf{return} $C_k$ \Comment{set of candidate $k$-itemsets}
\EndFunction
\end{algorithmic}
\end{algorithm}

\subsubsection{Subset determination}
\label{chap:subset_determination}
Finally it should be ensured that the comparison of frequent itemset candidates and
transactions in the database is evaluated efficiently.
For this, the candidates are stored in a hash-tree where each node refers to either a set of candidate itemsets (leaf)
or another node (inner node).
The depth $d$ of the node corresponds then to the hash of the $d$th item in the candidate itemset.
By recursively descending the hash tree for every suffix of a transaction (remainder) $t$,
a set of candidate itemsets is reached for each of which is checked whether it is contained in $t$.
If so, it is added to the answer set.
If the itemset $I$ is contained in $t$, its first item is contained in $t$ too.
By hashing on every suffix, all items in $t$ are first items once too, so there must occur
a match before missing any items.
After each descent, only the remaining items need to be considered.

\subsection{Discovering Rules from frequent itemsets}
\label{chap:discovering_rules}

As the confidence $conf(X \rightarrow Y)$ can be seen as equivalent to $Pr_D [Y|X]$,
$conf(X\rightarrow Y)$ is computed by dividing ${support(X\cup Y)}$ by ${support(X)}$.
When the support of each itemset is stored in the itemset generation process, this computation can be done quickly.
Still the number of association rules that can be extracted from each frequent itemset may be quite large.

Naively to discover all rules holding in a frequent itemset $I$,
all of the subsets $s \subset I$ would have to be evaluated whether the
rule $s \rightarrow (I - s)$ has minimum confidence.
If this is done for all frequent itemsets, the rule $s \rightarrow t, t \subseteq (I-s)$
is also checked as $s \cup t \subseteq I$ is also a frequent itemset.

A lot of confidence tests can be pruned.
First, 
\begin{equation*}
    \{s \rightarrow (X - s)| s \subset X\} = \{(X - s) \rightarrow s | s \subset X\}
\end{equation*}
Using the similarity to probability, it follows that
\begin{equation*}
    conf(X \rightarrow Y) = \frac{support(X\cup Y)}{support(X)} = \frac{support(X)}{support(X - s)}
\end{equation*}
If $\tilde{s} \subset s$ is inserted instead of $s$, it can be seen that $support(X-\tilde{s})$
decreases as $|X-s| < |X-\hat{s}|$. Thus the confidence of the rule increases.
Thus if $(X-s) \rightarrow s$ does hold, all $(X-\tilde{s}) \rightarrow \tilde{s}$ must also hold.
Like in \autoref{chap:apriori} the combination of rules with sufficient confidence can now be used
to generate candidate rules with larger consequents.

\begin{algorithm}
\caption{Apriori association rule generation from large $k$-itemsets and sets of $m$-item 
confidence satisfying consequents based on \cite{DBLP:conf/vldb/AgrawalS94}}
\begin{algorithmic}
\Procedure{as-rule-extraction}{frequent itemsets $L$}
  \ForAll{$L_k \in L$}
    \State $H_1 \gets \{ \{Y\} | Y \in L_k\}$
    \State \textbf{call} ap-gen-rules($L_k, H_1$)
  \EndFor
\EndProcedure
\Procedure{ap-gen-rules}{$l_k, H_m$}
  \If{$k > m$}
    \ForAll{$h_{m} \in H_{m}$}
      \State $c \gets$ support($l_k$)$/$support($l_k-h_{m}$)
      \If{$c \geq minconf$}
        \State \textbf{output} $(l_k - h_{m}) \rightarrow h_{m}$, conf: $c$, supp: support($l_k$)
      \Else
        \State $H_{m} \gets H_{m} - \{h_{m}\}$
      \EndIf
    \EndFor
    \State $H_{m+1} \gets$ apriori-gen($H_m$)
    \State ap-gen-rules($l_k, H_{m+1}$)
  \EndIf
\EndProcedure
\end{algorithmic}
\end{algorithm}

\subsection{Example}

\begin{figure}
    \centering
    \begin{tabular}{r|c|c|c|c|c}
        ID & A & B & C & D & E \\ \hline
        0  & 0 & 0 & 0 & 1 & 0 \\
        1  & 1 & 1 & 1 & 0 & 0 \\
        2  & 1 & 0 & 1 & 0 & 0 \\
        3  & 1 & 0 & 0 & 1 & 0 \\
        4  & 1 & 1 & 1 & 1 & 1 \\
    \end{tabular}
    \caption{Example transaction database for a market providing Aubergines, Beer, Charcoal, Dijon mustard and Edam cheese.}
    \label{fig:ex-trans-db}
\end{figure}

The apriori algorithm is shortly demonstrated based on the database shown in \autoref{fig:ex-trans-db} with
the attributes $Aubergine$, $Beer$, $Charcoal$, $Dijon\ mustard$ and $Edam\ cheese$.
Assume the user requests all association rules with minimum support of 30\% and minimum confidence of 60\%.
For the initialization, the set of frequent 1-itemsets is generated.
Only one transaction involves Edam. With a support of $\frac{1}{5}$, below the specified minimum support,
$\{Edam\}$ is not included in the set $L_1 = \{\{Aubergine\},\{Beer\},\{Charcoal\},\{Dijon\}\}$.
From this set, the new set of candidate itemsets of size 2 is generated.
As there are only sets of one item so far and no excluded items that could accidentally have been joined in,
$C_2$ is simply the cross product of the above set with itself.
Next, the whole database is scanned to compute the actual support of the generated candidates.
It turns out that $\{Charcoal, Dijon\}$ and $\{Beer, Dijon\}$ are too rare combinations (support of $\frac{1}{5}$) but all
remaining candidates satisfy the support condition.

The second iteration follows where $k = 3$ and sets of size 3 are generated from $L_2$.
For this, for example $\{A, C\}$ and $\{A, D\}$ can be joined to form $\{A, C, D\}$,
while $\{A, C\}$ and $\{C, D\}$ are not joined because their first elements already differ
 \footnote{And in this case also because $\{C, D\}$ is not a frequent itemset.}.
In the newly generated set we can still check for every element whether any of its subsets are non-frequent, which does mean
that we can prune it.
This is the case as we have not accepted $\{C, D\}$ in the previous iteration. $\{A, C, D\}$ is pruned from the candidate set.
After checking all valid combinations and ensuring the subset closure,
we retreive $\{Aubergine, Beer, Charcoal\}$ as the only candidate of size 3.
After a single scan of the database, we can ensure that it has support of $\frac{2}{5}$ and is accepted as frequent itemset.
The overall frequent itemsets are now all of the determined frequent itemsets of
all lengths.

The next step is the generation of association rules from the set of frequent itemsets.
The procedure will be shown by the example of the frequent itemset $\{A, B, C\} \in L_3$.
First, single consequent rules are generated and their confidence is computed, $AB \rightarrow C, c=\frac{2}{2}$, $AC \rightarrow B, c=\frac{2}{3}$ and $BC \rightarrow A, c=\frac{2}{2}$.
By coincidence all of the rules are accepted.
The new set of 2-item consequents is generated from the consequents forming $H_1$,
being (compare itemset generation) all pairs of items from $H_1$.
By computing the confidence for each rule, we retrieve $C \rightarrow AB, c=\frac{2}{3}$ and $B \rightarrow AC, c=\frac{2}{2}$ but delete
$A \rightarrow BC, c=\frac{2}{4}$.
In the next iteration the procedure stops as $k=3 \leq 3=m$.

\subsection{Interestingness Measures}

As can be seen in the above example, even from small itemsets, large amounts of association rules can be extracted.
Meanwhile there may be quite a few redundant rules among.
Knowing that $C \rightarrow AB$ and $B \rightarrow AC$, it might not be surprising that $CB \rightarrow A$.
Based on this, several interestingness measures have been proposed for pruning association rules
from the set of generated rules.
The urgency of filtered association rules becomes even more obvious when considering the case presented by Brin et al. in \cite{DBLP:conf/sigmod/BrinMS97}:
It is easy to construct cases where due to a large overall support of an item,
even negative correlations suffice to generate an association rule with.
This is demonstrated with \autoref{fig:neg-correlation},
where the rule $Tea \rightarrow Coffee$ is generated with a support of $25\%$ and confidence of
$\frac{25}{30} \approx 83\%$ which is quite high.
Yet, when considering that the probability of any customer drinking coffee is $90\%$ it can be seen that
this actually means a negative correlation between coffee and tea.

\begin{figure}
    \centering
    \begin{tabular}{r|c|c}
           & $c$ & $\bar{c}$ \\ \hline
        $t$  & 25 & 5 \\
        $\bar{t}$  & 65 & 5 \\

    \end{tabular}
    \caption{Example for small shop selling tea ($t$) and coffee ($c$) where the association rule $Tea\rightarrow Coffee$ with negative correlation is generated.}
    \label{fig:neg-correlation}
\end{figure}

Interestingness measures can be based on the expected value of an extracted rule thus redundancy or surprisingness,
as well as on utility or actionability. Further methods are possible as for the above example a chi-squared measure is proposed.
A detailed overview is presented by Geng and Hamilton in \cite{DBLP:journals/csur/GengH06}.

Another more theoretical way is presented by Pasquier et al. in \cite{DBLP:conf/icdt/PasquierBTL99} where from the reduced set of closures of itemsets
(the maximal set that has the same support as its subset) reduced association rules
are generated from which all original association rules could be generated but which also already serves a human understandable set of less redundant rules.

\section{Quantitative Association Rules}

Association rules consider only whether a product was bought or not.
Quantitative attributes like amount or price are not at all considered.
Still there might be of relations, for example
$Beer [\geq 3\text{ l}] \rightarrow Charcoal$.
This would express that large amounts of beer imply a grilling party.
Alternatively, one can imagine the number of seconds customers spend in front of the shelf
to be incorporated in the database.

The dataset of i.e. shopping transactions is now extended to include not only whether a specific product was bought,
but it also contains an associated quantity i.e. the amount of products bought.
Rules taking into account these quantities and especially all of their subranges can be
mined using the generic boolean association rule algorithm.
For this, ranges of quantities are introduced in place of every quantitative attribute
and for each item we store for each generated range whether the quantitiy item lied inside of the range or not.
If the dataset contains for example transactions including 1, 2 or 5 litre beer,
and this quantity was stored in the attribute beer before
we introduce the boolean attributes $[1,1], [2,2], [5,5], [1,2], [2,5] \text{ and } [1,5]$ corresponding to each interval.
If 2 litres beer were bought, $[2,2], [1,2], [2,5] \text{ and } [1,5]$ are now items in the transaction.
It can easily be seen that if all subranges are included, a quadratic amount of ranges is generated.
Even when restricting to ranges that the actual value lies included in,
there are on average $O(n^2)$ ranges that include a specific value \cite{DBLP:conf/sigmod/SrikantA96}.

If too few subranges are included it might happen that intervals that satisfy minimum support and confidence are excluded.
When restricted to equally sized intervals, choosing slim intervals,
the support for each interval could be too low.
In contrast, if the intervals are too wide, the confidence might be reduced \cite{DBLP:conf/sigmod/SrikantA96}.
At last, if an association rule containing a subrange does have minimum support,
all contained ranges do have minimum support, drastically increasing computation time.
Thus it should be carefully decided which ranges to include.

\subsection{Formal Definition}

In addition to \autoref{chap:formal_crisp_ass_rules} we define for each itemset $I$
a function $q_I: A \rightarrow N_0$ assigning each item in the set its quantity.
The quantity interval of attribute $a$ is defined as $\{ x \in N_0 | \exists{I \in D}: q_I(a) = x\}$

\subsection{Proposed Algorithm}

The algorithm prosposed by Srikant and Agrawal in \cite{DBLP:conf/sigmod/SrikantA96} introduces a user defined
maximum support and decomposes the transformation as follows:
\begin{itemize}
    \item Determine the number of partitions for each quantity interval (see \autoref{chap:quant_partitioning})
    \item Map the values in each quantitative interval to consecutive integers such that the order of the values is preserved.
    \item Find the support for each value of quantitative attributes and combine adjacent values that satisfy minimum support
            if they do not exceed the specified maximum support.
    \item Transform the itemset into boolean itemsets by replacing all quantitative attributes with the determined ranges.
\end{itemize}

After this procedure, the standard algorithm from \autoref{chap:apriori} is applied to generate boolean association rules.
In order to remove redundant rules regarding subintervals,
interestingness measures can again be introduced.

\subsection{Optimal Interval Partitioning}
\label{chap:quant_partitioning}

In order to measure the optimality of the interval partitioning,
so called "$K$-partial-completeness" is introduced in \cite{DBLP:conf/sigmod/SrikantA96}.
The intuitive idea is that for each rule $R$ that would be obtained when considering
all of the ranges over the involved quantitative attributes,
the generalized rule obtained by only considering the partitioned intervals
should be as "close" to $R$ as defined by the $K$.
"Closeness" is defined by having at most $K$ times the support of the rule $R$.
Essentially, every rule obtained by the partitioning should contain as few other quantities in the dataset as possible.

As Srikant et al.\cite{DBLP:conf/sigmod/SrikantA96} have shown, the number of required partitions is
\begin{equation*}
    N = \frac{2*N}{m*(K-1)}
\end{equation*}
Assuming that each partition equally splits the support, a partitioning in $N$ intervals of equal size
is generated.

\begin{figure}
    \centering
    \includegraphics[width=0.95\linewidth]{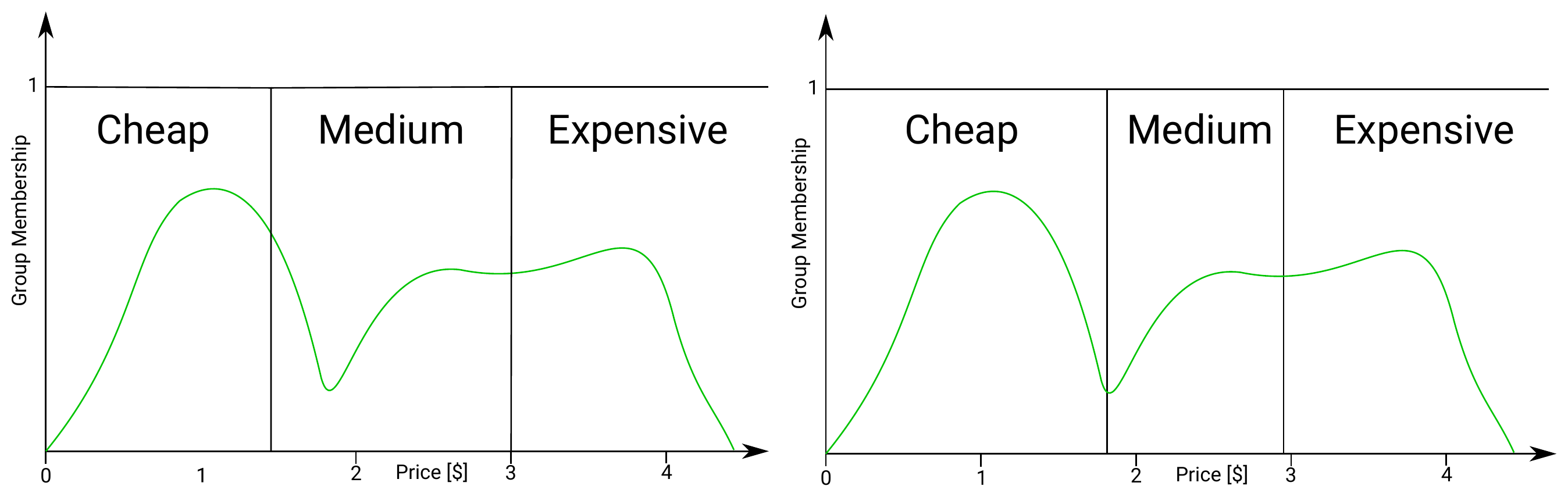}
    \caption{Graphs showing why sensititivity to the underlying data may be useful,
    based on approximate value distribution among the data in green.
    On the left a graph with equi-width subintervals is shown while on the right 
    subintervals were chosen based on the distribution of the data (clustering).}
    \label{fig:equi_width_vs_cluster}
    
\end{figure}

This assumption must not always hold as seen in \autoref{fig:equi_width_vs_cluster} which is why 
the intervals can be generated sensitive to the data by diverse clustering approaches.
As lots of these approaches are based on continuous values they are described at once.

\subsection{Continuous Intervals}

Considering continuous instead of discrete and finite quantitative attributes,
there is an infinite number of interval borders that can be chosen.
Alternatively to the equal-size approach, one can consider the available data when partitioning the interval.
A first approach is an equal frequency approach, where every partition contains the same amount of data points.
Advanced techniques apply a clustering of the feature interval,
trying to group along values with high frequency.
Examples for such procedures can be seen in \cite{preprocessing_numberical_data, 8601291}.


\subsection{Fuzzy Association Rules}

The before introduced concept of transforming ranges into
items can lead to problems at the border of an interval.
If i.e. $Beer [\geq 3\text{ l}] \rightarrow Charcoal$, is it so much more likely that a customer buys charcoal when
buying 3.0 litres beer instead of 2.9 litres?
A way to circumvent this is to make the importance or representativeness of values inside an interval decrease
with its proximity to the border of the interval.
This is generally achieved by introducing fuzzy sets.
Fuzzy value sets can overlap and have non-binary membership values.
A detailed introduction into fuzzy association rules by Helm can be found in \cite{MT:master_thesis/Helm07}.
As can be seen by comparing the work of Tan\cite{8601291} and Thomas, Raju\cite{DBLP:journals/itm/ThomasR14}, in both mining fuzzy rules
and quantitative rules, different clustering techniques are still a topic of high importance.

\section{Generalizing Association Rules}

\subsection{Motivation}

Consider again the database of the supermarket.
The manager of the supermarket might be interested in how to arrange
the items in the market such that all products from categories that are usually bought together
can be found in close shelves.

Until now a system can detect associations between specific products.
For the shelve problem, one would need a rule over the generalizations of products.
For example instead of $Charcoal, Aubergine \rightarrow Beer$, the rule
$Charcoal, Vegetables \rightarrow Beer$ might also hold.
At the same time, while $Aubergine \rightarrow Charcoal$ and $Courgette \rightarrow Charcoal$ might already hold,
the generalization $Vegetables \rightarrow Charcoal$ might not hold as the items are often bought together for barbecue
but don't make up the major part of vegetable concerned transactions. 
We will see that again as proposed by Srikant and Agrawl in \cite{DBLP:conf/vldb/SrikantA95}, the apriori algorithm can be used.
The procedure is described in the following subsections.

In addition to the dataset there now are taxonomies on the attributes of the database.
Instead of a forest of trees, these taxonomies are combined into a directed acyclic graph.

\subsection{Formal definition}

In addition to \autoref{chap:formal_crisp_ass_rules} a directed acyclic graph $T$
with all the items of the dataset as leafs is given, the taxonomy graph.
Item $x$ is called specialization of $\hat{x}$ and $\hat{x}$ is called generalization of $x$ if there
is an edge in $T$ from $\hat{x}$ to $x$.

\subsection{Basic Algorithm}

The support for the generalization of an attribute is not necessarily the sum of the supports
of its specializations.
This has the simple reason that one transaction can contain several specializations of the same item
and could already be seen in one of the motivating examples.
Hence a modification of the known apriori algorithm is necessary.

The most basic approach to this problem is to extend every transaction $t$ to a transaction $t'$, 
containing all the items and all of its ancestors.
For each item in the transaction, all of the ancestors are added.
Then with the algorithm from \autoref{chap:apriori},
association rules between these items can be extracted.
This algorithm works, but is quite inefficient.

Some simple optimizations proposed in \cite{DBLP:conf/vldb/SrikantA95} can directly be included.
First, when comparing transactions with candidate itemsets,
it is sufficient to include only those ancestors
that are element of any candidate itemset.
The generalizations of each item can be pre computed from $T$ at the beginning to save time.

A more sophisticated optimization is
pruning all itemsets that contain $x$ and $\hat{x}$.
The intuition is that, if a rule already contains an item,
adding its generalization trivially does not reduce its support.
More specifically it does not add any meaningful information,
so we can prune itemsets of that form (details can be found in \cite{DBLP:conf/vldb/SrikantA95}).


\subsection{Similarities to quantitative association rules}

\begin{figure}[h]
    \centering
    \includegraphics[width=0.5\linewidth]{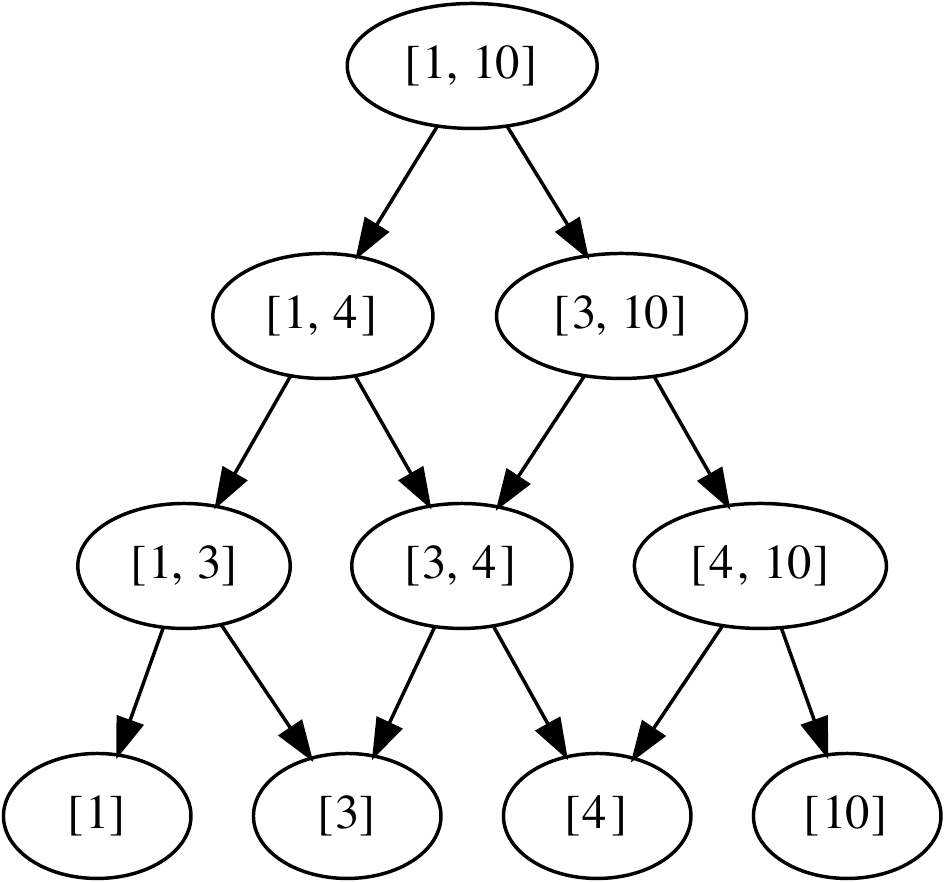}
    \label{fig:visualization-generalized-quantitative-q2g}
    \caption{Intervals of a quantitative attribute represented as a taxonomy.}
\end{figure}

When comparing quantitative and generalized association rules it might seem sensible
to conduct quantitative association rule mining by transforming the quantitative attributes to items in a taxonomy.
In the quantitative approach, an "optimal" partition of the overall interval is searched for
such that only the partition intervals have to be considered.
Contrasting this with the generalization approach, a multi-level subinterval approach emerges by introducing each superinterval as a generalization
of its sub interval in the taxonomy tree.
This of course is technically equivalent to considering each possible superinterval for a value of an itemset.
As already noted by Srikant et al. \cite{DBLP:conf/sigmod/SrikantA96}, each value lies in $O(n^2)$ subintervals when there are $n$ distinct values for the attribute.
For few values of the quantitative attribute this may be useful as it avoids loss of information.
Otherwise, the use of efficient pruning techniques becomes even more important.
Applying clustering to quantitative values may still be useful as a preprocessing step
for drastically reducing the size of the generated taxonomy tree.

\begin{figure}[h]
    \centering
    \includegraphics[width=0.5\linewidth]{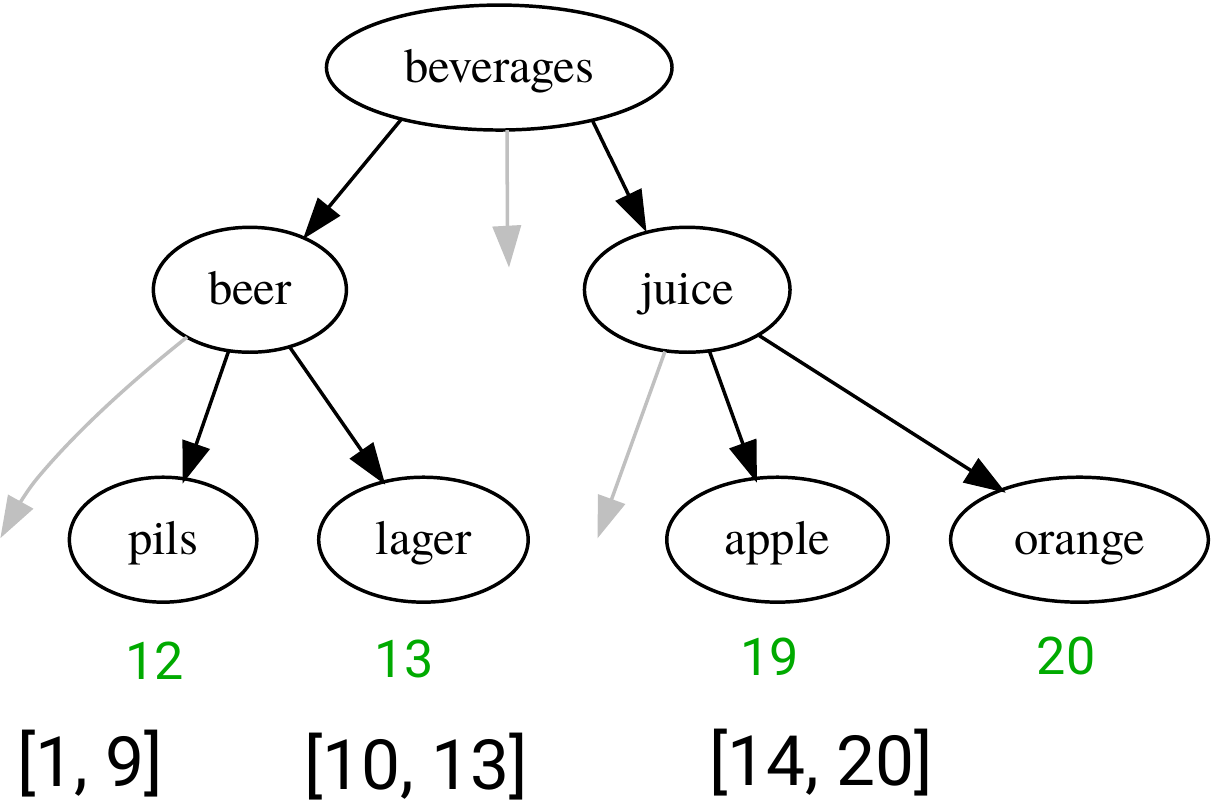}
    \label{fig:visualization-generalized-quantitative-g2q}
    \caption{A taxonomy converted into quantities. The subintervals shown below could be considered for association rules.}
\end{figure}

Similarly, in a tree-formed taxonomy, each leaf can be mapped to a number from left to right.
All different leaf level specializations are thus only regarded as a quantity of the top level generalization.
With this change, instead of considering items of every level,
only groups of items over the hierarchy are considered in the form of an interval.
Yet due to the free choice of interval borders, any lowest common ancestor can result in being considered,
regardless of its level in the taxonomy.
Clustering may then be used to reveal which ancestors are worth being considered most.
It has to be noted this is not applicable for multiple joined taxonomies that result in a non-tree-formed DAG.
This could make creating non-overlapping and monotonous intervals impossible.
The usefulness of this approach has to be evaluated for each case of application individually
as the handling of quantitative rules ensures a loss of information.
On the other hand, this approach is able to handle large taxonomies
\footnote{As described above only $O(n)$ intervals are considered instead of a number of $O(n^2)$ generalizations in the original quantitative approach}.

In both cases, advanced interestingness rules applied to each problem can be transferred to
either variation.

\section{Use cases}
Apart from providing information for market experts, association rules can also be used in recommender systems for new users
by recommending items that were frequently bought by others with similar shopping baskets \cite{DBLP:conf/pakdd/ShawXG10}.
Or for finding people that influence each other in social networks by finding associations between comments on posts \cite{DBLP:journals/entropy/ErlandssonBBJ16}.

Because of the set structure of association rules, they are not easily suitable for order dependent rule mining.
For tight associations and predictions of new data points as in interpolation,
association rules are not suitable.
For datasets $\{x, y=x^2\}$ it could be discovered that $X:High \rightarrow Y:High$
and $X:Low \rightarrow Y:Low$ using quantitative rules, but categories instead of values are associated.

\section{Summary and Outlook}

The apriori algorithm iteratively generates itemsets that increase in size step by step.
In the process, it prunes the evaluation of many association rules
by exploiting the downward closure property of support and confidence.
It can be seen that the rules extracted should be evaluated by domain experts
and at least checked against actual correlation before further usage. 
When applicable they can be used in many different domains,
especially market analysis.

For quantitative attributes,
the standard algorithm can be extended and efficiently improved by transforming ranges of values to single attributes.
In generalized association rules, all itemsets are extended by the ancestors of all contained items.
For non-diverse quantitative attributes or very large taxonomies it might even
be suitable to convert either extension into the other.

In the future, drawbacks and benefits of this interconversion may be evaluated.
The comparison can also be extended to include objective oriented
utility based association rules as introduced by Shen et al. in \cite{DBLP:conf/icdm/ShenZY02}.
Also, the extensions to allow for fuzzy sets and continuous intervals
pose further challenges regarding sensible set operations and efficient pruning
and provide interesting aspects to be evaluated in an overview.






\bibliographystyle{IEEEtran}

\bibliography{IEEEabrv,references}
%
%
%


\end{document}